\documentclass[doublespacing]{elsart}

 \usepackage{graphicx}

\usepackage{amssymb}
\usepackage{amsmath}

\begin{document}
\begin{frontmatter}



\title{A SPH model for incompressible turbulence}


\author{X. Y. Hu}
\author{and N. A. Adams}

\address{Lehrstuhl f\"{u}r Aerodynamik und Str\"{o}mungsmechanik, Technische Universit\"{a}t
M\"{u}nchen
\\ 85748 Garching, Germany}

\begin{abstract}
A coarse-grained particle model for incompressible Navier-Stokes (NS) equation 
is proposed based on spatial filtering by utilizing 
smoothed particle hydrodynamics (SPH) approximations.
This model is similar to our previous developed SPH discretization of NS equation
({\it Hu X.Y. \& N.A. Adams, J. Comput. Physics}, {\bf 227}:264-278,
2007 and {\bf 228}:2082-2091, 2009)
and the Lagrangian averaged NS (LANS-$\alpha$) turbulence model.
Other than using smoothing approaches, 
this model obtains particle transport velocity 
by imposing constant $\sigma$ which is associated with the particle density, 
and is called SPH-$\sigma$ model.
Numerical tests on two-dimensional decay and forced turbulences with high Reynolds number 
suggest that the model is able to reproduce both the inverse energy cascade 
and direct enstrophy cascade of the kinetic energy spectrum,
the time scaling of enstrophy decay and the non-Guassian probability density function (PDF)    
of particle acceleration.
\end{abstract}
\end{frontmatter}
\section{Introduction}
The smoothed particle hydrodynamics (SPH) method is a fully Lagrangian, grid free method. 
Since its introduction by Lucy \cite{Lucy1977} and Gingold and Monaghan \cite{GingoldMonaghan1977}, 
SPH has been applied to a wide range of macroscopic and mesoscopic flow problems \cite{Monaghan2005}\cite{HuAdams2006}. 
Though the SPH method was originally developed for astrophysical problems involving compressible fluids,
it has been extended to problems involving incompressible fluids by using either a weakly compressible model of the 
fluid \cite{Monaghan1994}, 
or by algorithms designed to solve the full incompressible equations 
\cite{CumminsRudman1999}\cite{hu2007imp}\cite{hu2009constant}.

Many of the incompressible flow problems, such as flood and coastal flows, to which SPH has been applied are turbulent. Since the direct numerical simulation of these problems is not always feasible, turbulence modeling is required for the computational more efficient coarse-grained numerical simulation.
One straightforward approach of SPH turbulence modeling is applied the turbulence models originally developed for Euelrian methods directly
\cite{gotoh2003sph}\cite{shao2006incompressible}\cite{violeau2007numerical}. 

Monaghan \cite{monaghan2002sph} first noticed the similarity between the version of SPH called XSPH \cite{monaghan1989problem}
and the Lagrangian averaged Navier-Stokes (LANS) turbulence model \cite{holm1999fluctuation}\cite{holm2002averaged} 
on the relation between the velocity determined from particle momentum 
(momentum velocity) and the transport velocity,
and proposed a turbulence model specifically for the SPH method.
In this model, the SPH particle moves with the transport velocity
smoothed from momentum velocity by an iterative algorithm 
and a dissipation term is introduced to mimic 
the standard large eddy simulation (LES) model originally developed for Eulerian methods.
A further modification of this model (SPH-$\epsilon$) is to obtain transport velocity directly by the XSPH method with a parameter $\epsilon$ \cite{monaghan2009turbulence}.
On the other hand, we have noticed the importance of SPH particle moving with the
velocity different from its momentum velocity when simulating flows beyond small Reynolds number
in our previous developed incompressible SPH method \cite{hu2007imp}\cite{hu2009constant}.
In this method, other than using the XSPH method or smoothing approaches, 
the Eulerian incompressibility condition (zero velocity divergence)
and the Lagrangian incompressibility condition (constant density) 
are used respectively to determine the momentum velocity and the transport velocity. 

In this paper, we propose a coarse-grained particle system for turbulence simulation 
based on spatial filtering the Navier-Stokes (NS) equation by utilizing SPH approximations. 
Since the resulting particle equations are similar to those of the above mentioned incompressible SPH method,
except for an additional effective stress term introduced by moving particle with transport velocity,
the same numerical method is applied. 
The numerical tests show that, while achieving good accuracy for resolved flow, 
the present model can recover the spectral and statistical properties 
of the two-dimensional decay and forced turbulences with high Reynolds number. 
\section{Model}
We consider the incompressible isothermal NS equation in Lagrangian form
\begin{eqnarray}\label{governing_equation}
\frac{d\mathbf{v}}{dt} & = & \frac{\partial \mathbf{v}}{\partial t} + \mathbf{v}\cdot \nabla \mathbf{v} = -\frac{1}{\rho}\nabla p +
\nu \nabla^2 \mathbf{v},
\label{equation_of_motion} \\
\frac{d\rho}{dt} & = & 0 \quad {\rm or} \quad \nabla \cdot \mathbf{v} = 0, \label{continuous-equation}
\end{eqnarray}
where $\rho$, $p$ and $\mathbf{v}$ are fluid density,
pressure and velocity, respectively, and $\nu =
\eta/\rho$ is the kinematic viscosity.
Note the two expressions
(constant density and zero velocity-divergence) in Eq.
(\ref{continuous-equation}) are formally equivalent.

The equation of the LANS-$\alpha$ model \cite{holm1999fluctuation}\cite{holm2002averaged} is a coarse-grained equation, 
written in the form of Eulerian mean or filtered velocity $\widetilde{\mathbf{v}}$
and Lagrangian mean or transport velocity $\widehat{\mathbf{v}}$ as
\begin{eqnarray}\label{governing_equation-na-s}
\frac{\partial\widetilde{\mathbf{v}}}{\partial t} + \widehat{\mathbf{v}}\cdot \nabla \widetilde{\mathbf{v}} & = & -\frac{1}{\rho}\nabla \widetilde{p} +
\nu \nabla^2 \widetilde{\mathbf{v}},
\label{equation_of_motion-ns-a} \\
\nabla \cdot \widehat{\mathbf{v}} & = & \nabla \cdot \widetilde{\mathbf{v}} = 0, \label{continuous-equation-ns-a} \\
\widetilde{\mathbf{v}} & = & (1 - \alpha^2\nabla^2)  \widehat{\mathbf{v}}. \label{ns-a}
\end{eqnarray}
The Helmholtz operator $(1 - \alpha^2\nabla^2)$ in Eq. (\ref{ns-a}) suggests that the transport velocity 
$\widehat{\mathbf{v}}$ is smoother than the filtered velocity $\widetilde{\mathbf{v}}$.
One can think of the parameter $\alpha$ as the length scale associated with 
the width of an extra filter which smooths $\widetilde{\mathbf{v}}$ to obtain $\widehat{\mathbf{v}}$.
This has been done explicitly in the SPH-$\epsilon$ model \cite{monaghan2009turbulence}, 
where $\widehat{\mathbf{v}}$ is based on the filtering 
\begin{equation}
\widehat{\mathbf{v}}  = \widetilde{\mathbf{v}} + \epsilon \int \widetilde{\mathbf{v}}(\mathbf{r}') -\widetilde{\mathbf{v}}(\mathbf{r}) G(\mathbf{r}' - \mathbf{r} , l)d\mathbf{r}', 
\label{filtering-monaghan}
\end{equation}
where $G$ is the filter with width $l$ and $\epsilon$ is a constant parameter.
The basic idea of smoothing is to prevent the production of small length-scale flow structures.

In the next section, we propose a coarse-grained particle system 
based on filtering the NS equation with numerical techniques in the SPH method. 
This approach is different from the SPH-$\epsilon$ model,
which is devised from the SPH discretized form of Eckart's Lagrangian \cite{eckart1960variation}.
\subsection{Coarse-grained NS equation and SPH method}
Assume that the incompressible flow field is coarse-grained into a particle system with spatial filtering,
the variables on particles are obtained by
\begin{equation}
\psi_i  =  G_i * \psi  =  \int \psi W(\mathbf{r} - \mathbf{r}_i, h)d\mathbf{r}, 
\label{filtering}
\end{equation}
where $\mathbf{r}_{i}$ and $h$ are the center and width, respectively, of the filter,
$W(\mathbf{r} - \mathbf{r}_{i}, h)$, which is radially symmetric respect to $\mathbf{r}_{i}$ and
has the properties $\int W(\mathbf{r}-\mathbf{r}_{i}, h)
d\mathbf{r} =1 $, $\int\nabla W(\mathbf{r} - \mathbf{r}_i)d\mathbf{r} = \mathbf{0}$ 
and $\lim_{h\rightarrow
0}W(\mathbf{r}-\mathbf{r}_{i}, h) = \delta(\mathbf{r}-\mathbf{r}_{i})$.
Note that this filter can also take the role as the SPH kernel function with smoothing length $h$. 
Substitute $\psi$ with the coordinate $\mathbf{r}$ into Eq. (\ref{filtering}),
one has the particle position $\mathbf{r}_i$ at the center of the filter.
The motion of particle is determined by its transport velocity, i.e.
\begin{equation}\label{filtered-particle-position}
\frac{d \mathbf{r}_i} {dt} =  \widehat{\mathbf{v}}_i.
\end{equation}
Note that $\widehat{\mathbf{v}}_i$ can be different from the filtered or momentum particle velocity $\mathbf{v}_i$,
which is obtained by substitute $\psi$ with velocity $\mathbf{v}$ into Eq. (\ref{filtering}).
Similarly, the filtered derivatives on particles can be written as
\begin{eqnarray}
G_i * \nabla \psi & = & \int \nabla\psi W(\mathbf{r} - \mathbf{r}_i, h)d\mathbf{r} \nonumber \\
& = & - \int \psi \nabla W(\mathbf{r} - \mathbf{r}_i, h)d\mathbf{r}, \label{derivative} \\
G_i * \frac{d \psi}{dt} & = & \frac{\partial \psi_i}{\partial t}  + \int \nabla\psi\cdot \mathbf{v} W(\mathbf{r} - \mathbf{r}_i, h)d\mathbf{r} \nonumber \\
& = & \frac{\partial \psi_i}{\partial t} - \int \psi \mathbf{v}\cdot \nabla W(\mathbf{r} - \mathbf{r}_i, h)d\mathbf{r}, \label{full-derivative} 
\end{eqnarray}
Since only the filtered values $\psi_i$ are known, 
it is impossible to apply the exact filtering $G_i * \nabla \psi$ and $G_i * d\psi/d t$,
an approximated filtering is carried out after the flow field is first reconstructed 
\begin{equation}
\psi  \approx  \frac{\sum_j\psi_j W(\mathbf{r} - \mathbf{r}_j)}{\sum_k W(\mathbf{r} - \mathbf{r}_k)}
 = \frac{1}{\sigma}\sum_j\psi_j W(\mathbf{r} - \mathbf{r}_j),
\label{reconstruction} 
\end{equation}
where $\sigma = \sum_k W(\mathbf{r} - \mathbf{r}_k)$ 
is a measure of local length scale,
which is larger in a dense particle region than in a
dilute particle region \cite{KoshizukaNobeOka1998}.
From the identity $\sum_j W(\mathbf{r} - \mathbf{r}_j)/\sigma = 1$,
the total volume $V$ can be written as
\begin{equation}\label{volume}
V = \sum_j V_j = \sum_j \int\frac{1}{\sigma}W(\mathbf{r} - \mathbf{r}_j)d\mathbf{r} \approx \sum_j \frac{1}{\sigma_j}, 
\end{equation}
where $\sigma_j$ equals to the inverse of particle volume approximately, i.e. $\sigma_j = \sum_k W(\mathbf{r}_j - \mathbf{r}_k) = \sum_k W_{jk} \approx 1/V_j$.
With Eq. (\ref{derivative}) and by using Eq. (\ref{reconstruction}) and the properties of the kernel function, 
the approximation of $G_i * \nabla \psi$ can be obtained by
\begin{eqnarray}
G_i *\nabla \psi  & \approx & - \sum_j \overline{\psi}_{ij}  \int\frac{2}{\sigma}\nabla W(\mathbf{r} - \mathbf{r}_i)W(\mathbf{r} - \mathbf{r}_j)d\mathbf{r} \approx  \sum_j \frac{2}{\sigma_j} \overline{\psi}_{ij} \nabla W_{ij}, \label{approx-derivative}
\end{eqnarray}
where $\overline{\psi}_{ij} = (\psi_{i} + \psi_{j})/2$ is the average between particle $i$ and particle $j$. 
Note that Eq. (\ref{approx-derivative}) is a typical SPH discretization of the gradient operator.
Substitute $\psi$ with pressure $p$ and velocity $\mathbf{v}$, respectively, in Eq. (\ref{approx-derivative}), the filtered pressure gradient and velocity divergence
can be approximated as
\begin{equation}
G_i *\nabla p \approx \sum_j \frac{2}{\sigma_j} \overline{p}_{ij} \nabla W_{ij}, 
\quad G_i *\nabla \cdot \mathbf{v} \approx \sum_j \frac{2}{\sigma_j} \overline{\mathbf{v}}_{ij} \nabla W_{ij}.
\label{pressure-gradient}
\end{equation}
Similarly, if the average directional derivative is approximated as 
$\overline{\nabla \psi}_{ij} \approx \mathbf{e}_{ij}(\psi_{i} - \psi_{j})/r_{ij}$, 
where $\mathbf{e}_{ij}$ and $r_{ij}$ are unit vector and distance, respectively, from particle $i$ to particle $j$,
the filtered velocity Laplacian can be approximated with Eq. (\ref{approx-derivative}) as
\begin{equation}
\quad G_i *\nabla^2 \mathbf{v} = G_i *\nabla \cdot \nabla\mathbf{v}  \approx  \sum_j \frac{2}{\sigma_j} \frac{\mathbf{v}_{ij}}{r_{ij}} \mathbf{e}_{ij}\cdot\nabla W_{ij}.
\label{velocity-laplacian}
\end{equation}
Using Eq. (\ref{volume}), the particle density is given by
\begin{equation}\label{density_summation}
\rho_{j} = m_{j}\sigma_{j},
\end{equation}
where $m_i$ is the constant mass of a particle.
The variation of the particle density is 
\begin{equation}
\frac{d\rho_{i}}{dt}= -  \rho_{i}\sum_j \frac{1}{\sigma_i} \widehat{\mathbf{v}}_{ij} \cdot  \nabla W_{ij},
 \label{density-variation}
\end{equation}
where $\widehat{\mathbf{v}}_{ij} = \widehat{\mathbf{v}}_{i} - \widehat{\mathbf{v}}_{j}$ is the difference of transport velocity between particle $i$ and particle $j$.
Using particle transport velocity $\widehat{\mathbf{v}}_{i}$, $G_i * d\psi/d t$ can be rewritten as
\begin{eqnarray}
G_i *\frac{d \psi}{dt} & = & \frac{\partial \psi_i}{\partial t} - \int \psi \mathbf{v}\cdot \nabla W(\mathbf{r} - \mathbf{r}_i h)d\mathbf{r} \nonumber \\
& = & \frac{\partial \psi_i}{\partial t} + \widehat{\mathbf{v}}_{i}\cdot \nabla \psi_i 
+ \int \psi (\widehat{\mathbf{v}}_{i} - \mathbf{v})\cdot \nabla W(\mathbf{r} - \mathbf{r}_i, h)d\mathbf{r} \nonumber \\
& \approx & \frac{d \psi_i}{d t} +   \sum_j \frac{1}{\sigma_j} \psi_{ij} \widehat{\mathbf{v}}_{ij}\cdot \nabla W_{ij},
\label{approx-full-derivative}
\end{eqnarray}
where $\psi_{ij} = \psi_{i} - \psi_{j}$. 
Substitute $\psi$ with momentum density $\rho \mathbf{v}$ in Eq. (\ref{approx-full-derivative}), one has
\begin{equation}
G_i *\frac{d \rho \mathbf{v}}{dt}   \approx   \frac{d \rho_i \mathbf{v}_i}{d t} + \sum_j \frac{1}{\sigma_j} 
(\rho \mathbf{v})_{ij} \widehat{\mathbf{v}}_{ij}\cdot \nabla W_{ij}, 
\label{approx-momentum}
\end{equation}
where $(\rho \mathbf{v})_{ij} = \rho_i \mathbf{v}_i - \rho_j \mathbf{v}_j$, 
and the second term is an additional effective stress term.
With Eqs. (\ref{derivative}), (\ref{density-variation}) and (\ref{approx-momentum}), 
Eqs. (\ref{equation_of_motion}) and (\ref{continuous-equation}) can be coarse-grained into 
\begin{eqnarray}\label{governing_equation-filtered}
m_{i}\frac{d \mathbf{v}_i}{d t} & = & -\sum_j \frac{2}{\sigma_i\sigma_j} \left[\overline{p}_{ij} \nabla W_{ij}
+ \frac{1}{2} (\rho \mathbf{v})_{ij} \widehat{\mathbf{v}}_{ij}\cdot \nabla W_{ij}
- \eta\frac{\mathbf{v}_{ij}}{r_{ij}}\mathbf{e}_{ij}\cdot \nabla W_{ij} \right],
\label{equation_of_motion-filtered} \\
& & \sum_j \frac{1}{\sigma_i} \widehat{\mathbf{v}}_{ij} \cdot  \nabla W_{ij}  =  \sum_j \frac{2}{\sigma_j} \overline{\mathbf{v}}_{ij} \cdot \nabla W_{ij} = 0. \label{continuous-equation-filtered}
\end{eqnarray}
The particle system of Eqs. (\ref{filtered-particle-position}), (\ref{density-variation}), (\ref{equation_of_motion-filtered}) and (\ref{continuous-equation-filtered}) is very similar to 
the SPH discretization of NS equation used in Hu and Adams \cite{hu2007imp}\cite{hu2009constant}, 
except that there is an additional effective stress term in Eq. (\ref{equation_of_motion-filtered}).
Actually, without the additional effective stress term,
Eqs. (\ref{equation_of_motion-filtered}) 
and (\ref{continuous-equation-filtered}) 
are a SPH discretization of the Eqs. (\ref{equation_of_motion-ns-a}) 
and (\ref{continuous-equation-ns-a}) of the LANS-$\alpha$ model.
Since this term is induced by filtering, its contribution is expected to be small 
when the flow is well resolved as the influence of filtering is negligible.
Note that this term is similar to the additional effective stress term in 
the SPH-$\epsilon$ model, where only the filtered velocities are token into account together 
with the tunable parameter $\epsilon$. 

To solve this coarse-grained system,
the transport velocity $\widehat{\mathbf{v}}_{i}$ is obtained from 
the filtered velocity $\mathbf{v}_{i}$.
For the present particle system, 
since the local length scale is characterized by $\sigma_i$
which variates according to $\widehat{\mathbf{v}}_{i}$, 
preventing the production of small length scale
is equivalent to constrain the constant density condition.
This is already included in the present system 
as the first expression of Eq. (\ref{continuous-equation-filtered}).
Similar to the denotation of the SPH-$\epsilon$ model, 
in which $\widehat{\mathbf{v}}_{i}$ is obtained
by a smoothing approach with a parameter $\epsilon$,
the present  particle system is denoted as the SPH-$\sigma$ model since $\widehat{\mathbf{v}}_{i}$ 
is obtained by constraining $\sigma_i$.
\section{Numerical method} 
As mentioned above that the presented particle system is very similar to 
the SPH discretization of NS equation used in Hu and Adams \cite{hu2007imp, hu2009constant}, 
the same fractional time-step integration approach can be used. 
First, the transport velocity  or half-time-step velocity is
obtained by
\begin{equation}\label{particle-velocity}
\widehat{\mathbf{v}}_{i} =  \mathbf{v}^{n}_{i} + \left(\mathbf{f}^{n}_{i}
-\frac{1}{\rho}\nabla p'_{i}\right)\frac{\Delta t}{2}.
\end{equation}
where $\Delta t$ is time-step size and $\mathbf{f}$ is the sum of the viscous term and the additional effective stress term. 
The pressure field $p'_{i}$ is to impose the constant density condition,
i.e. the first expression in Eq. (\ref{continuous-equation-filtered}).
Subsequently, the particle position at the new time
step is calculated by
\begin{equation}
\mathbf{r}^{n+1}_{i}  =  \mathbf{r}^{n}_{i} + \widehat{\mathbf{v}}_{i}
\Delta t \label{particle-position},
\end{equation}
and the particle velocity at the new time step is obtained by
\begin{equation}
\mathbf{v}^{n+1}_{i}  =   \mathbf{v}^{n}_{i}  + \left(\mathbf{f}^{n}_{i}
-\frac{1}{\rho}\nabla p_{i}\right) \Delta t,
\label{full-step-velocity}
\end{equation}
where the pressure field $p_{i}$ is to impose the
zero velocity-divergence, i.e. the second expression in Eq. (\ref{continuous-equation-filtered}).

\subsection{Constant-density condition}
We split Eq.(\ref{particle-velocity}) into an intermediate step and
a correction step. An intermediate velocity
$\widehat{\mathbf{v}}^{*}_{i}$ is obtained by
\begin{equation}
\widehat{\mathbf{v}}^{*}_{i}  =  \mathbf{v}^{n}_{i} + \frac{\Delta
t}{2}\mathbf{f}_{i}\left(\mathbf{r}^{n},\mathbf{v}^{n} \right),
\label{first-intermediate}
\end{equation}
and the intermediate pressure is obtained by solving the discretized Poisson equation
\begin{equation}
 \nabla \cdot \left( \frac{\nabla
p^{*}}{\rho}\right)^{n}_{i} = - \frac{\Delta t}{\sigma_{i}}\sum_{j} \nabla W_{ij} \cdot
\widehat{\mathbf{v}}^{*}_{ij} + \frac{\sigma^{0}_{i} - \sigma^{n}_{i}}{\sigma^{n}_{i}}.
\label{Poisson-new}
\end{equation}
where $p^{*} = \frac{1}{2}p' \Delta t^{2}$, $\sigma^{0}_{i}$ is the initial value
and the left-hand-side is
\begin{equation}
\nabla \cdot \left(\frac{\nabla p^{*}}{\rho}\right)_{i}  =  -
\frac{1}{\sigma_{i}}\sum_{j} \nabla W_{ij}\cdot
\left[\frac{1}{m_{i}} \sum_{k} \frac{2}{\sigma_{i}\sigma_{k}} \overline{p}^{*}_{ik} \nabla
W_{ik} -
\frac{1}{m_{j}}\sum_{l} \frac{2}{\sigma_{j}\sigma_{l}} \overline{p}^{*}_{jl} \nabla
W_{jl}\right].
 \label{exact-projection}
\end{equation}
Note that Eq. (\ref{Poisson-new}) takes account the accumulated density error,
which is given by the second term on the right-hand-side. 
The generalized minimum residual (GMRES ($q$), $q$ is the number of
inner iteration steps without restarting) method is used to solve Eq. (\ref{Poisson-new})
with the convergence condition $\min(Re_i)  <  \epsilon$, where $Re_i$ is residue,
suggesting that the maximum relative variation of $\sigma_{i}$ is about $\epsilon$, typically from 1\% to 0.1\%. 
With Eq. (\ref{particle-position}), 
the new particle position can be obtained by
\begin{equation}\label{discrete-pressure-2}
\mathbf{r}^{n+1}_{i}  =  \mathbf{r}^{n}_{i} + \Delta t
\widehat{\mathbf{v}}^{*}_{i} - \frac{1}{m_{i}}\sum_{j} \frac{2}{\sigma_{i}\sigma_{j}} \overline{p}^{*}_{ij} \nabla
W_{ij}.
\end{equation}
\subsection{Zero velocity-divergence condition}
An intermediate velocity at the full time step $\mathbf{v}^{*,
n+1}$ is obtained by
\begin{equation}
\mathbf{v}^{*, n+1}_{i}  =  \mathbf{v}^{n}_{i}
+\Delta t\mathbf{f}_{i}\left(\mathbf{r}^{n},\mathbf{v}^{n}
\right). \label{inter-velocity}
\end{equation}
The pressure is obtained by solving the discretized Poisson equation
\begin{equation}
\sum_j \frac{2}{\sigma_j} \frac{p^{**}_{ij}}{\overline{\rho}_{ij}r_{ij}}\mathbf{e}_{ij}\cdot \nabla W_{ij} =  \sum_j \frac{2}{\sigma_j} \overline{\mathbf{v}}^{*, n+1}_{ij} \cdot \nabla W_{ij}
\label{Poisson},
\end{equation}
where $p^{**} = p \Delta t$. Note that, different from the exact projection for the constant-density constraint,
the zero-velocity-divergence constraint is enforced by approximate projection, see details in Hu and Adams \cite{hu2007imp}\cite{hu2009constant}.
The velocity at the full time step $\mathbf{v}^{n+1}$ is obtained
by
\begin{equation}
\mathbf{v}^{n+1}_{i}  =  \mathbf{v}^{*, n+1}_{i} - \frac{1}{m_{i}}\sum_{j} \frac{2}{\sigma_{i}\sigma_{j}} \overline{p}^{**}_{ij}.
\label{new-velocity}
\end{equation}
\subsection{Time-step criteria}
Besides viscous-diffusion condition,
different from Hu and Adams \cite{hu2007imp}\cite{hu2009constant},
in the present work, the so-called Lipshitz  
or deformational CFL (DeCFL) condition is used as the stability condition
\begin{equation}\label{time-step}
\Delta t \leq 0.25 \min\left(\frac{0.5}{||\nabla \mathbf{v}_i||_{\max}},\frac{h^2}{\nu}\right).
\end{equation}
where $||\nabla \mathbf{v}_i||_{\max}$
gives the magnitude of the maximum strain rate of the flow.
The DeCFL condition suggests that the trajectories of the particles 
can not cross each other. 
Numerical experiments show that the DeCFL condition allows larger time-step size than the CFL condition 
if the flow is moderate. 
However, for flow with high shear rate, the two conditions gives comparable time-step sizes. 
Also note that, though large time-step size according to 
Eq. (\ref{time-step}) does not lead to numerical instability,
it increases the temporal truncation error.
\section{Numerical simulations}
Several two-dimensional test cases are provided to assess the potential of the SPH-$\sigma$ model. 
First, the performance of the model for recovering flow with moderate Reynolds number is tested. 
Then two-dimensional decay and driven turbulent flows with high Reynolds number are tested.
In order to achieve sufficient high Reynolds number, the physical viscosity in NS equation is switched off.
For all cases a quintic spline kernel \cite{MorrisFoxZhu1997} is used as smoothing
function. A constant smoothing length $h$, which is kept equal to the
initial distance between the neighboring particles, is used for all
test cases. As elliptic solver a diagonally
preconditioned GMRES ($q$) method is used. 
For all test cases,  
the computation is performed on a domain $0<x<1$ and $0<y<1$ with
periodic boundary conditions in both directions. 
For spectrum analysis with fast Fourier transform (FFT), 
the values on a uniform grid are reconstruct from the particles by 
a remesh method with a 4th order interpolation kernel \cite{ChaniotisPoulikakosKoumoutsakos2002},
which is much more computational efficient than the
SPH Fourier transform used in Robinson and Monaghan \cite{robinson2011direct}.
\subsection{Two-dimensional Taylor-Green flow}
The two-dimensional viscous Taylor-Green flow is a periodic array of
vortices, where the velocity
\begin{equation}
u =  -Ue^{bt}\cos(2\pi x)\sin(2\pi y), \quad
v =  Ue^{bt}\sin(2\pi x)\cos(2\pi y) 
\label{tg-theory}
\end{equation}
is an exact solution of the incompressible NS equation.
$b=-8\pi^2/{\rm Re}$ is the decay rate of velocity field. We
consider a case with $\rm Re = 100$, which has been used to test
different incompressible SPH methods
\cite{ChaniotisPoulikakosKoumoutsakos2002}\cite{ellero2007isp}\cite{hu2007imp}\cite{hu2009constant}.
The computational setup is the same as that of \cite{hu2009constant}. 
The initial particle velocity is assigned according to Eq. (\ref{tg-theory}) by
setting $t=0$ and $U=1$. Same as in \cite{hu2009constant}, the initial
configuration with 3600 particle is taken from previously stored particle
positions (relaxed configuration).

\begin{figure}[p]
\begin{center}
\includegraphics[width=1.2\textwidth]{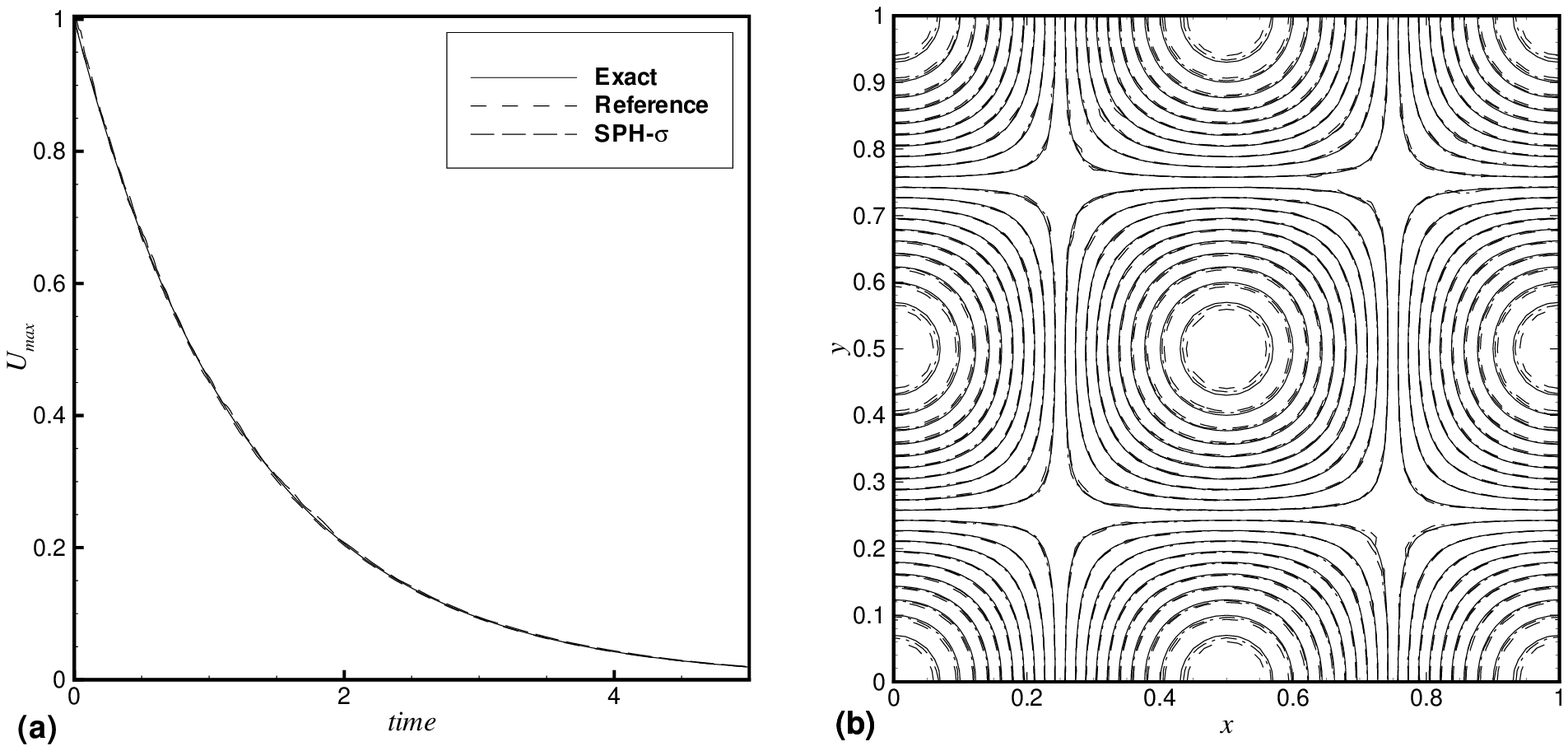}
\end{center}
\caption[]{Taylor-Green problem simulated with 3600 particles: 
(a) decay of the maximum velocity (reference: solution in \cite{hu2009constant}), 
(b) vorticity field (dashed
line: solution in \cite{hu2009constant}, dash-dot line: SPH-$\sigma$) and
exact solution (solid line).}
\label{tgv}
\end{figure}

Figure \ref{tgv} shows the calculated decay of the maximum velocity
and the vorticity field at $t=1$.
It can be observed that the present solution is in quite good
agreement with the exact and reference solutions \cite{hu2009constant}.
This is not unexpected since the only notable difference between the SPH-$\sigma$ model
and the SPH discretization in \cite{hu2009constant} is the additional effective stress term 
in Eq. (\ref{equation_of_motion-filtered}), 
whose contribution is small since the flow is well resolved.
Note that if  the additional effective stress term is not applied, 
as expected, the numerical solution (not shown here) has no noticeable difference 
with the reference solution. Note that, compared to the reference solution, 
the present solution is slightly less dissipative,
suggested by the somewhat less errors in regions close to the centers of vortex cells.
This behavior suggests that, 
as will also be shown in the next case, 
the overall effect of the additional effective stress term is decreasing dissipation.
\subsection{Two dimensional decay turbulence}
We consider a two dimensional decay turbulence with high Reynolds number.
The simulations are carried out with two resolutions, i.e. with $50\times 50$ and $100 \times 100$ particles.
The particles were placed initially on a grid of squares 
in motion with velocities specified by a $8 \times 8$ array of
Taylor-Green vortices
\begin{equation}
u  =  \cos(8\pi x)\sin(8\pi y), \quad
v =  \sin(8\pi x)\cos(8\pi y).
\label{tg-decay}
\end{equation}
This initial condition mimics the experimental setup 
in which arrays of vortices are generated by applying electric field \cite{paret1998intermittency}, 
or movable solid bars \cite{maassen2002self}. 

\begin{figure}[p]
\begin{center}
\includegraphics[width=1.2\textwidth]{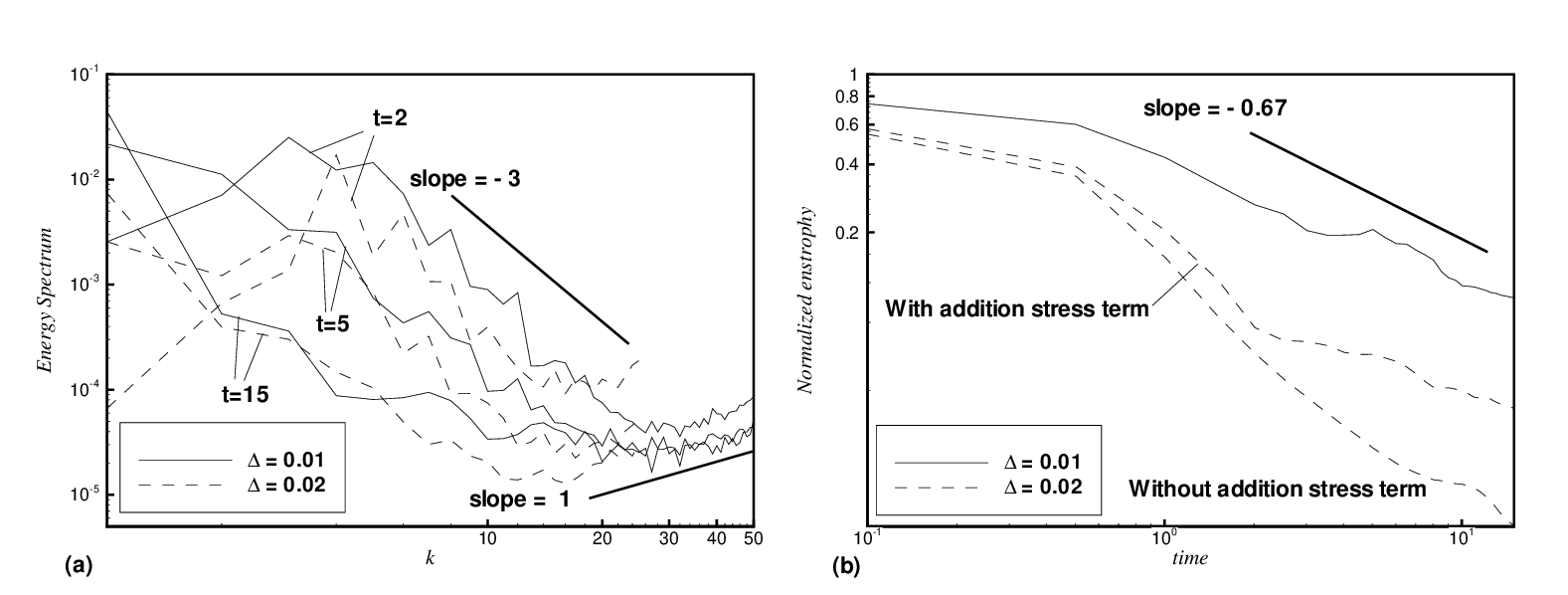}
\end{center}
\caption{Two dimensional decay turbulence: 
(a) energy spectrum at $t=2$, 5 and 15. 
(b) evolution of total kinetic energy and enstrophy.
The computations are performed on the periodic domain of $[0, 0]\times[1, 1]$
with $50\times 50$ and $100\times 100$ particles.} \label{decay1}
\end{figure}

\begin{figure}[p]
\begin{center}
\includegraphics[width=1.2\textwidth]{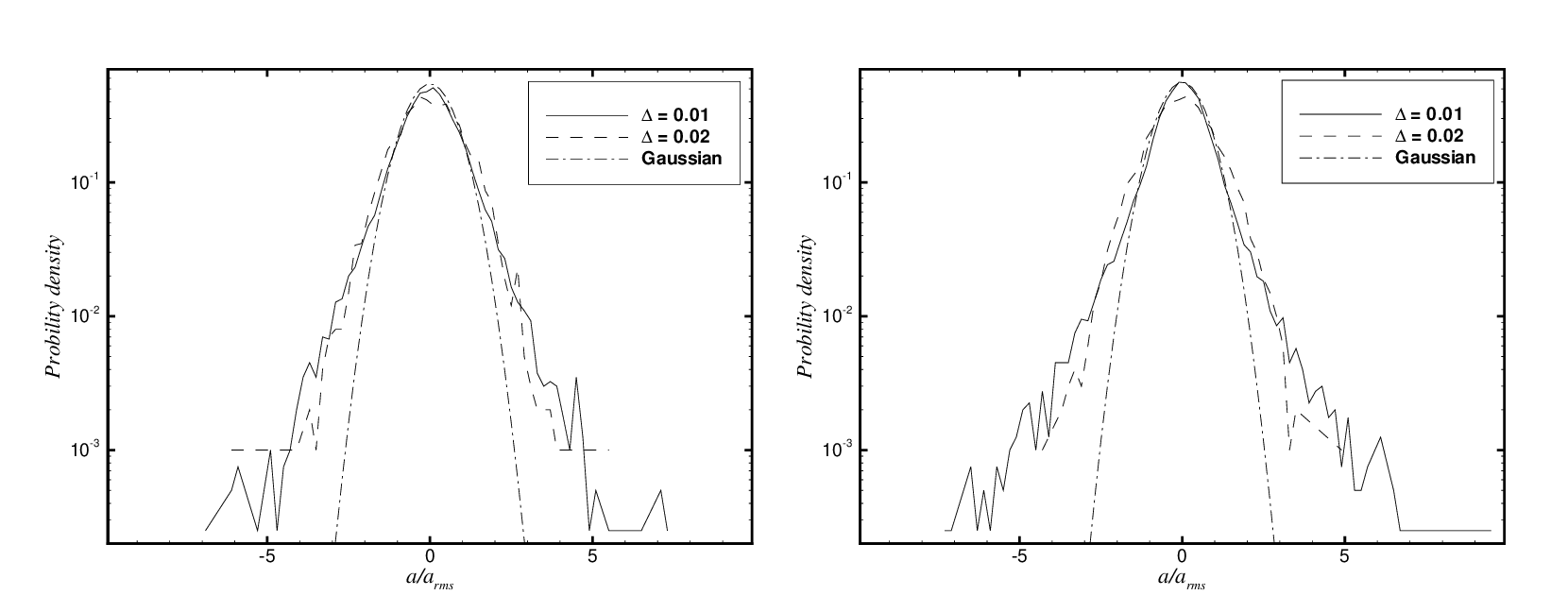}
\end{center}
\caption{Two dimensional decay turbulence: PDF of acceleration at (a) $t=2$ and (b) $t=15$.} 
\label{increment}
\end{figure}

The energy spectrum at $t = 2$, 5 and 15, and the evolution of the normalized total enstrophy are given in Fig. \ref{decay1}. 
As shown in Fig. \ref{decay1}a, 
full turbulence has been developed at $t=2$, 
and is characterized by a typical direct entropy cascade 
with $-3$ scaling of the energy spectrum \cite{tabeling2002two} 
independent of the resolution of simulation.
The spectrum at high wave numbers is that of white noise.
This reflects the SPH method's limitation on resolving flow structured 
at small scales close to the smoothing length.
Similar phenomenon is also observed in experiments, such as in \cite{paret1999vorticity},
where the measured spectrum is dominated by white noise at the high wave numbers beyond resolvability.
As shown in Fig. \ref{decay1}b, the overall decay of enstrophy scales at about $t^{-0.5} - t^{-0.7}$,
independent of the resolution of simulation,
which is in good agreement with previous direct numerical simulation \cite{chasnov1997decay} 
and theoretical prediction \cite{yakhot2004universal}. 
Note that if the additional effective stress term in Eq. (\ref{equation_of_motion-filtered}) is not included in the model, 
as shown in Fig. \ref{decay1}b, the enstrophy decay is considerably faster in the low resolution simulation. 
\begin{figure}[p]
\begin{center}
\includegraphics[width=1.2\textwidth]{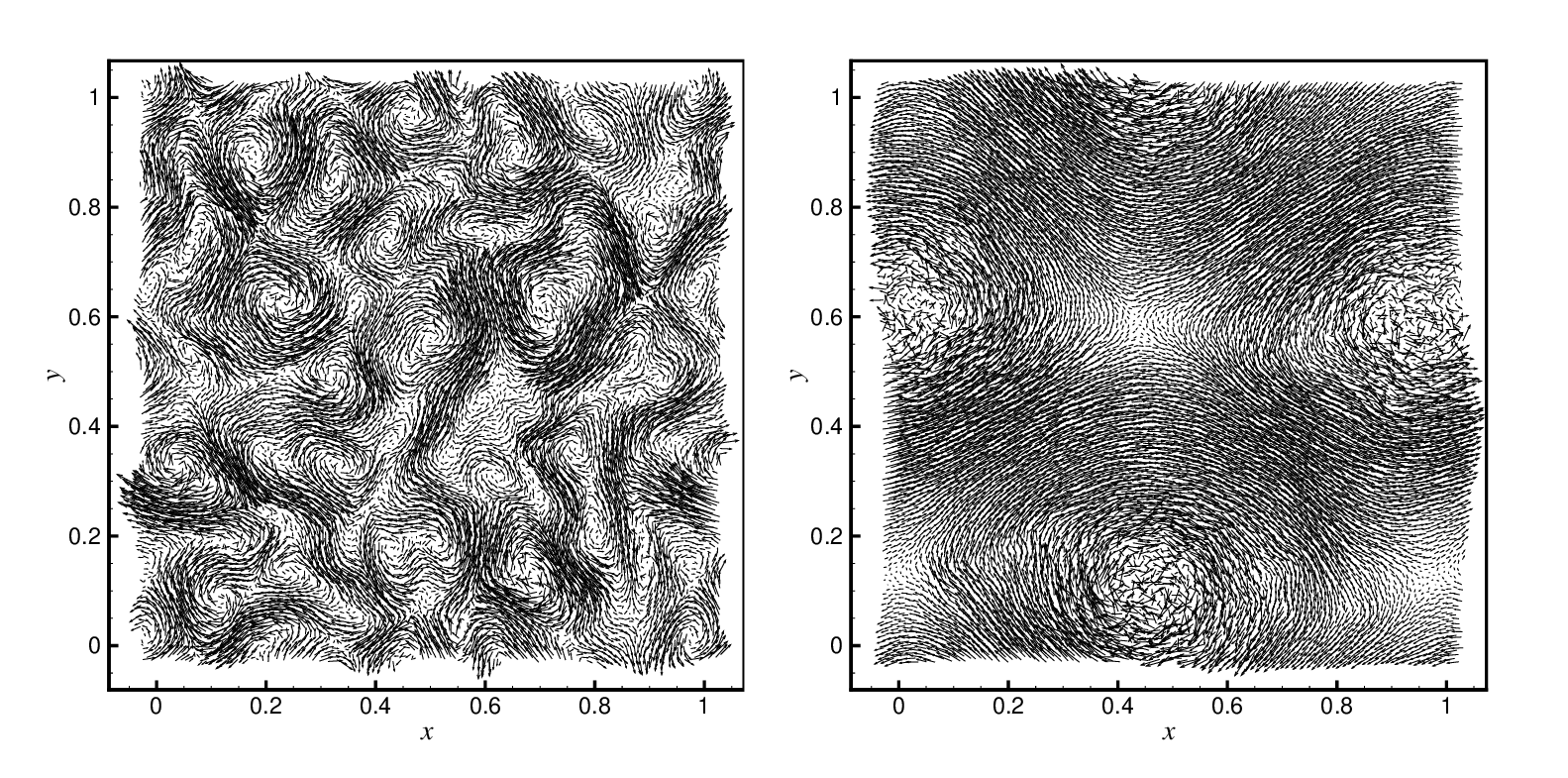}
\end{center}
\caption{Two dimensional decay turbulence: velocity field at (a) $t=2$ and (b) $t=15$.} \label{decay2}
\end{figure}
One important property of two-dimensional decay turbulence is the merging and pairing of vortices.
As the velocity fields at $t=2$ and 15 shown in Fig. \ref{decay2}, the initially smaller vortices finally develop to the configuration with a pair of big vortices comparable to square-size with opposite sign, 
which is known as the equilibrium state of two-dimensional decay turbulence \cite{tabeling2002two}.
This change is also reflected in the energy spectrum, as shown in Fig. \ref{decay1}, 
where the energy spectrum is considerably flatten.
The probability density function (PDF) of velocity increments along the particle trajectory or acceleration at $t = 2$ and 15 is shown Fig. \ref{increment}. 
While small acceleration fluctuations follow a Gaussian distribution,
large acceleration fluctuations show a distinct non-Gaussian. 
This PDF suggests the intermittency of the velocity field, which is
well established for turbulence.
Note that, while the turbulence is approaching the equilibrium state,
other than disappear, the intensity of intermittency increases slightly.
\subsection{Two dimensional forced turbulence}
We consider a two dimensional forced turbulence with high Reynolds number.
The simulation is carried out with $100 \times 100$ particles,
which were placed initially on a grid of squares with zero velocity.
The driving force is pairwise and produced by an inverse-viscous term
\begin{equation}
\mathbf{f}^{drv}_{ij} = -\frac{2}{\sigma_i\sigma_j} 
\eta^{drv}\frac{\mathbf{v}^{drv}_{ij}}{r_{ij}}\mathbf{e}_{ij}\cdot \nabla W_{ij},
\label{tg-decay}
\end{equation}
where $\eta^{drv} = 2.5 \times 10^{-3}$ and the velocity profile $\mathbf{v}^{drv}$ is obtained from
the equation
\begin{equation}
u^{drv} =  \cos(16\pi x + \phi)\sin(16\pi y + \phi), \quad
v^{drv} =  \sin(16\pi x + \phi)\cos(16\pi y + \phi),
\label{tg-decay}
\end{equation}
where the phase shift $\phi$ is produced with increments of a Wiener process.
Again, this driving force mimics the experimental setup 
which drives an arrays of vortices by applying electric field 
\cite{paret1998intermittency}. 
The difference is that zero mean flow is achieved strictly in the simulation.

\begin{figure}[p]
\begin{center}
\includegraphics[width=1.2\textwidth]{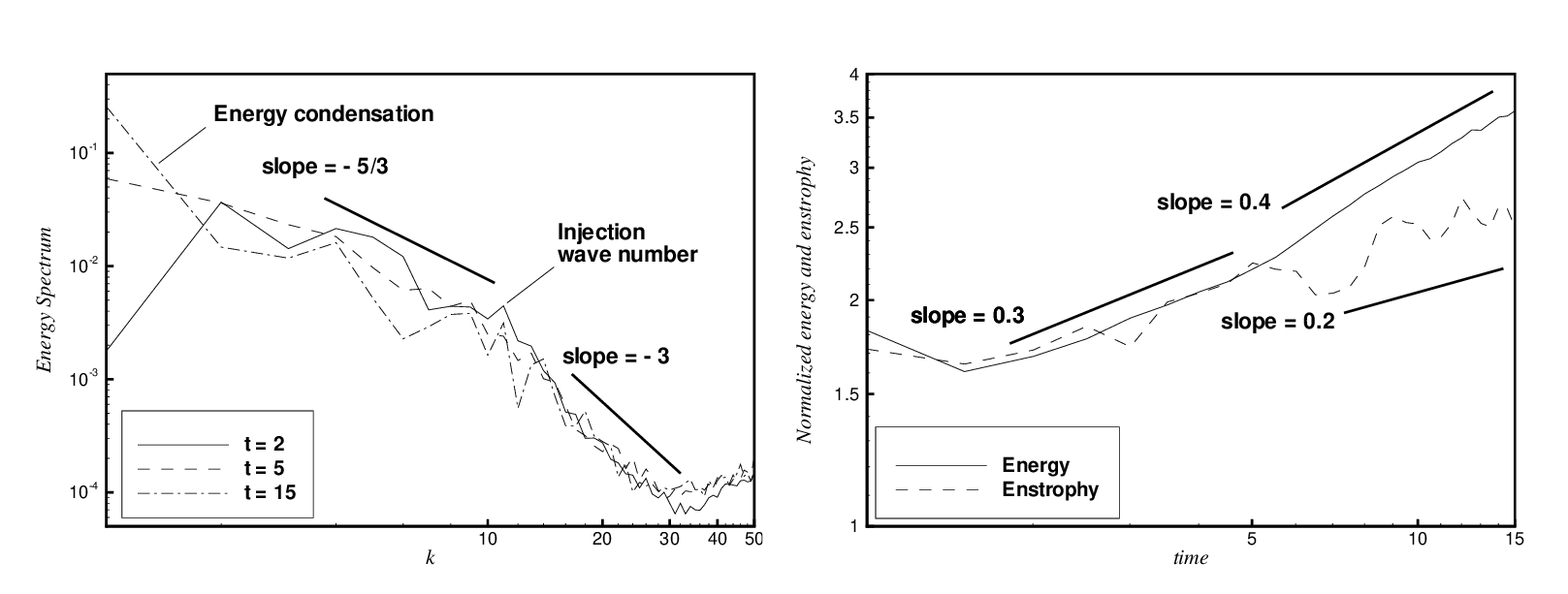}
\end{center}
\caption{Two dimensional forced turbulence: (a) energy spectrum at $t=2$, 5 and 15. 
(b) evolution of total kinetic energy and total enstrophy.
The computations are performed on the periodic domain of $[0, 0]\times[1, 1]$
with $100\times 100$ particles.} \label{driving}
\end{figure}

The energy spectrum at $t = 2$, 5 and 15, and the evolution of the total energy and total enstrophy (both are normalized with the values at $t=0.5$) are given in Fig. \ref{driving}. 
As shown in Fig. \ref{driving}a, 
full turbulence has been developed at $t=2$,
in which depending on smaller or larger than the injection wave number,
the energy spectrum show a direct entropy cascade 
with $-3$ scaling and an inverse energy cascade, 
which suggests transporting kinetic energy to larger length scales, 
with $-5/3$ scaling \cite{tabeling2002two}.
As there is energy injection, 
the kinetic energy and enstrophy increase at $t^{0.3}$ before about $t=5$.
After this time, while the enstrophy increase is slightly slowed down to about $t^{0.2}$,
the kinetic energy increase is slightly sped up to $t^{0.4}$. 
This change is corresponding to the time, also see in Fig. \ref{driving}a, 
from which the energy with the largest possible scale (square size) piles up.
As shown in Fig. \ref{driving}a, 
at $t=15$ the turbulence reaches the so-called condensed state, 
where the energy at the largest scale is considerable large than 
that represented by a $-5/3$ inverse-cascade scaling and still increases with time.
\begin{figure}[p]
\begin{center}
\includegraphics[width=1.2\textwidth]{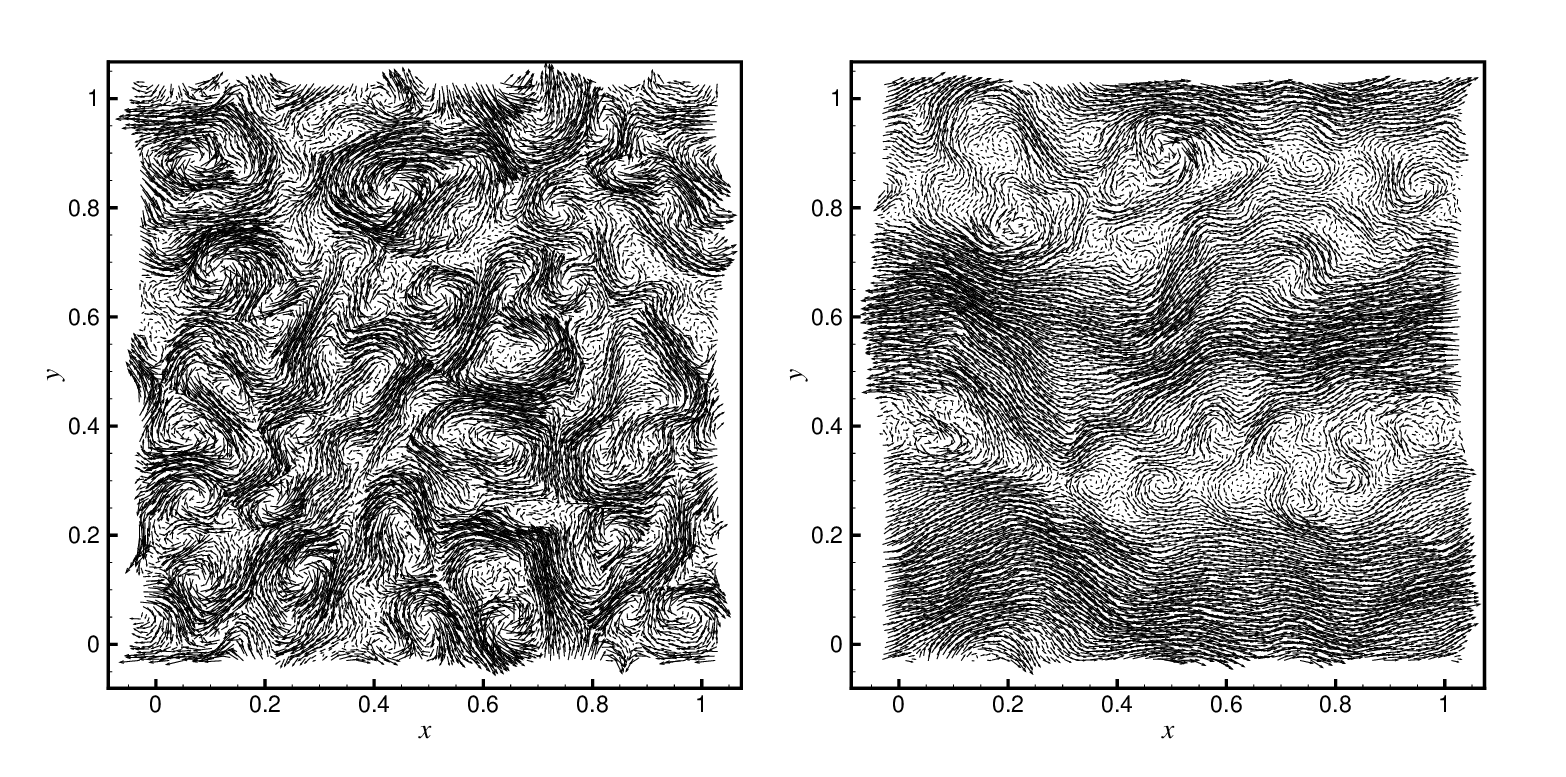}
\end{center}
\caption{Two dimensional forced turbulence: velocity field at (a) $t=2$ and (b) $t=15$.} \label{drv2}
\end{figure}
As shown in Fig. \ref{drv2}, 
while the velocity field at the early stage of forced turbulence resembles that of the decay turbulence,
it is very different at the late time, when the forced turbulence reaches the condensed state 
and the decay turbulence the equilibrium state, though both are dominated with large-scale flow structures.
In the condensed sate, different from the decay turbulence, there is no large vortex but large scale shearing flow.
In addition, there are small vorticies corresponding the injection wave number, 
but no small scale vorticies found in the equilibrium sate of a decay turbulence, 
as shown in Fig. \ref{decay2}b.
\begin{figure}[p]
\begin{center}
\includegraphics[width=1.2\textwidth]{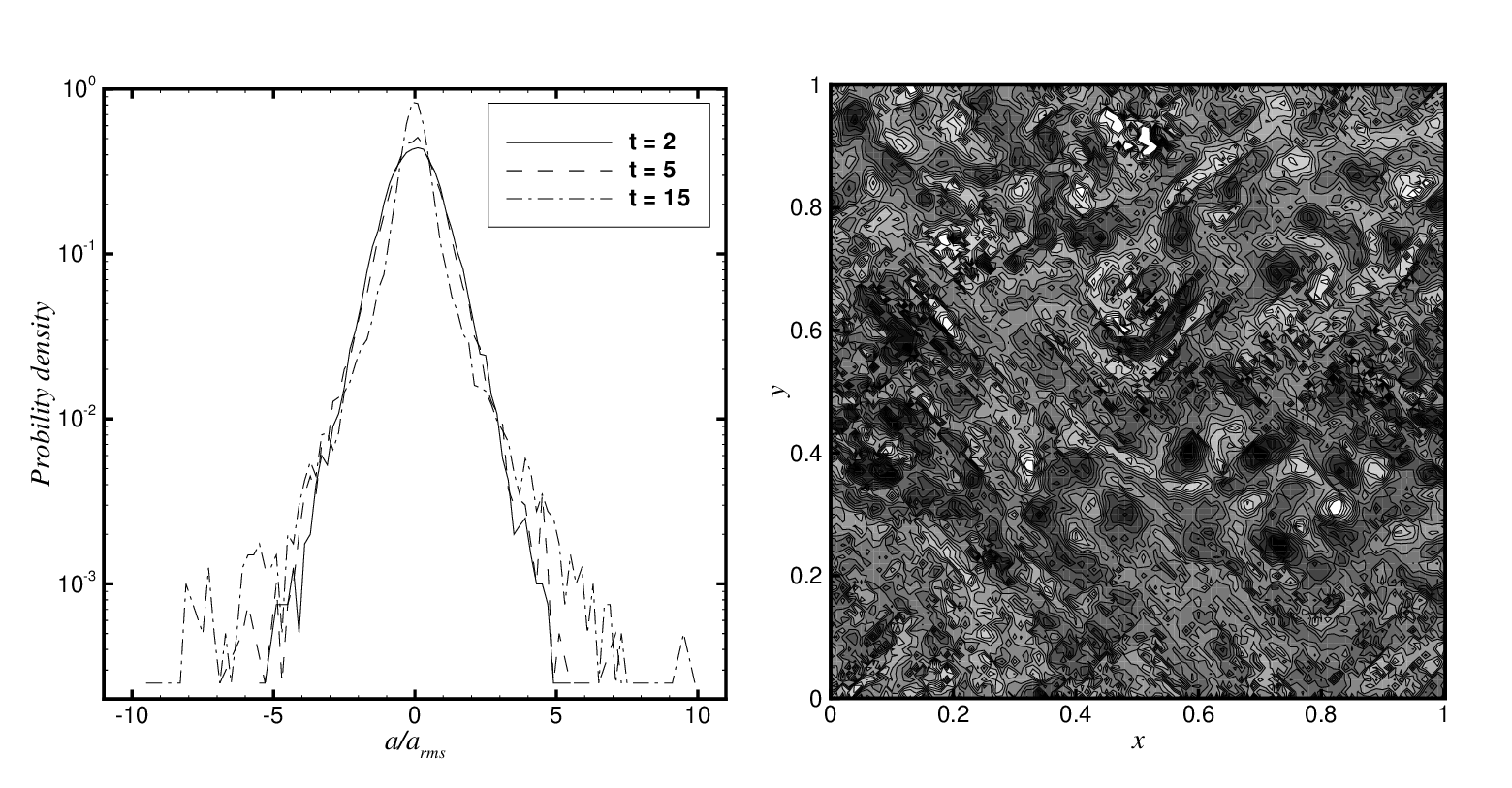}
\end{center}
\caption{Two dimensional forced turbulence: (a) PDF of acceleration at $t=2$, 5 and 15;
(b) 15 vorticity contours from $-50$ to $50$ at $t=15$.} \label{increment-drv}
\end{figure}
The PDF of acceleration at $t = 2$, 5 and 15 is shown Fig. \ref{increment-drv}a. 
It can be observed that before the condensed state, 
the PDF is in good agreement with that of decay turbulence before the equilibrium state.
Compared to the decay turbulence in the equilibrium state, 
the intensity of the intermittency increases considerably in the condensed state,
which is indicated by the narrowed probability for small fluctuation 
and extended probability for large fluctuation \cite{tabeling2002two}. 
This is also reflected in the vorticity contour of Fig. \ref{increment-drv}b,
where a number of spatially highly intermittent 
small pockets are presented with large positive and negative vorticity.
\section{Concluding remarks}
We have proposed a coarse-grained particle (SPH-$\sigma$) model for incompressible NS equation 
based on spatial filtering by utilizing SPH approximations.
The SPH-$\sigma$ particle model is similar to the LANS-$\alpha$ model 
and the SPH-$\epsilon$ model but with the different additional effective stress term
and approach to obtain the particle transport velocity.
Since, numerically, this model resemble to 
the discretization of our previous developed incompressible SPH method,
the same numerical method is applied.
Numerical tests on two-dimensional moderate flow, and decay and forced turbulences are carried out.
The results suggest that this model can be used for simulating 
incompressible turbulent flow problems to which SPH has been applied.
\bibliographystyle{plain}
\bibliography{../reference/SPH/sph-literature}
\end{document}